\documentstyle[12pt]{article}
\input{amssym.def}
\begin{document}

\begin{center}
{\bf \LARGE
\baselineskip=24pt

The Integrated Density of States for 1D Nanostructures at Zero
Bias Limit

\baselineskip=14pt

}

\vskip 1cm

{\large
L.A.Dmitrieva
}

\vskip1cm
Department of Mathematical and Computational Physics,
St.Petersburg State University,
198504 St.Petersburg, Russia\\
E-mail: mila@JK1454.spb.edu

\end{center}
\vskip1cm
\begin{center}
{\large Abstract}
\end{center}

By methods of quasiclassical asymptotics the behaviour of the
integrated density of states for 1D periodic nanostructures at
the zero bias limit is studied. It is shown that the density of
states at the zero bias limit has no regular limit while the
integrated density of states has. The rigorous proof of this
phenomenon given in the paper is based on a novel approach for
the quasiclassical asymptotics on the spectrum of the
Stark-Wannier operators. A connection of this phenomenon with
the zero bias limits of the current through the nanostructures
and their conductivity is briefly discussed.

\newcommand{\tg}{\mathop{\rm tg}\nolimits}
\newcommand{\tcnh}{\mathop{\rm tcnh}\nolimits}
\newcommand{\const}{\bf const}
\newcommand{\ctg}{\mathop{\rm ctg}\nolimits}
\newcommand{\RR}{\mbox{$\Bbb{ R}$}}
\newcommand{\ZZ}{\mbox{$\Bbb{ Z}$}}

\section{Introduction}

Some experimental works with tunneling devices have observed an increasing
conductivity near zero bias \cite{A1}. This "zero-bias anomaly" has been
explained in a number of papers by different physical phenomena \cite{A2,
A3, A4}. Some of these explanations are based on the electron density of
states and its behaviour near zero bias. In the present paper we study the
behaviour of the density of states and integrated density of states at
zero bias and show that the first one has no regular limit while the
latter has.

Because the current through the device and, in turn, the conductivity are
defined through the integral from the density of states and some
smooth function, it means that we should look for a "weak limit" of the
density of states. We proof rigorously that the integrated density of
states tends to an appropriate limit as the strength of the applied
electric field goes to zero. This prove is the main novelty  of the paper.

The main mathematical objects in our constructions are the spectral
projections kernels of the operator
\begin{equation}
H=-\frac{d^2}{dx^2} + p(x) -\varepsilon x,
\label{H}
\end{equation}
which describes 1D periodic structure  embedded in an external electric field with the
strength $\varepsilon > 0$ \footnote {It might be the 1D
periodic nanostructure generated by the stimulated Mott-Peierls
transition on quantum wire (see [6-10] )
with the bias applied} . The function $p$ is assumed to be a periodic one.
The operator $H$ has been an object of the intensive
investigations (see \cite{B1, B2, B3, B4, B5, G1} and references
therein). However
to our best knowledge the behaviour of spectral projections and of the
integrated density of states which is expressed in their terms have not
been considered as $\varepsilon \to 0$.

In papers \cite{B3, B4, B5} there has been performed the quasiclassical
study of the equation $Hg=Eg$. These results have been used in
\cite{B1,B2} for the solution of a number of spectral problems related
to the operator $H$. For instance, there has been studied the behaviour
of the residues of the analytical continuation of the resolvent kernel of
$H$ and proved that in an appropriate sense the ladders of resonances are
concentrated $\varepsilon \to 0$ on the spectral band of the operator
$H_0$. However the methods used in the above papers do not allow
to study the behaviour of the operator $H$ characteristics on
the spectrum which fills the whole axis. In the present paper we
overcome the difficulties related to the asymptotics
constructions at real energies and hence provide the
mathematical basis for studying the integrated density of
states and the conductivity  near the zero bias limit.

One of the advantages of the approach proposed in the present
paper is that we explicitly take into account in the leading
order of $\varepsilon$ the
tunneling probabilities through spectral gaps for the device
embedded into the electric field. These probabilities are
known \cite{B4, B5, G1} to be exponentially small w.r.t. to
$\varepsilon$ however they strongly influence the behaviour of the
integrated density of states.

\section{Preliminaries}

In this section we recall some facts concerning the one-dimension
Schr\"{o}dinger operator
$$
H_0=-\frac{d^2}{dx^2} + p(x)
$$
with a periodic potential $p(x+a)=p(x)$ and the Hamiltonian $H$
given by (\ref{H}).

The spectrum of $H_0$ in $L_2(\RR )$ consists of bands
$[\lambda_{2l-2}, \lambda_{2l-1}]$ separated by gaps $(\lambda_{2l},
\lambda_{2l-1})$ \cite{T}. In the present paper we assume that the number of gaps
in the spectrum is finite, i.~e. $l=1,2,\ldots, N$. Let us note that the
class of the $N$-gap potentials has been described completely \cite{FG} in
connection with the algebraic-geometrical integration of the
Korteveg-de Vries equation. However in the present paper we do not need
the knowledge of the explicit form of these potentials but only make use
of the fact that number of gaps is finite.

The integrated density of states for the operator $H_0$ is given by the
formula \cite{A5}:
\begin{equation}
N(\lambda )=\frac{1}{a}\int\limits_0^a e(x,x,\lambda)dx,
\label{DS0}
\end{equation}
where $e(x,y,\lambda)$ is the spectral function of $H_0$:
$$
e(x,y,\lambda)=\sum_{j=1}^{m}e_j(x,y)
$$
if $\lambda$ belongs to the $m+1$-gap $(\lambda_{2m}, \lambda_{2m+1})$ and
$$
e(x,y,\lambda)=\sum_{j=1}^{m}e_j(x,y) +
\frac{1}{\pi}\int\limits_{\lambda_{2m}}^{\lambda}
\frac{\Re [\psi(x,\lambda)\overline{\psi(y,\lambda)}]}{\dot
 {\lambda}(k)}d\lambda
$$
if $\lambda$ belongs to the $(m+1)$-th band $[\lambda_{2m},
 \lambda_{2m+1}]$. Here $e_j(x,y)$ is the kernel of the spectral
 projection of $H_0$ on the $j$-th band  $\Delta_j=[\lambda_{2j-2},
 \lambda_{2j-1}]$:
\begin{equation}
e_j(x,y)=\frac{1}{\pi}\int\limits_{\Delta_j}
\frac{\Re [\psi(x,\lambda)\overline{\psi(y,\lambda)}]}{\dot
 {\lambda}(k)}d\lambda.
\label{KH0}
\end{equation}
The Bloch functions $\psi(x,\lambda)$ here are assumed to be normalized by the
condition
\begin{equation}
\frac{1}{a}\int\limits_{0}^{a}
\psi (x, \lambda )\overline{\psi (y, \lambda)}dx=1.
\label{N}
\end{equation}
By $\dot {\lambda}(k)$ we denote $\frac{d}{dk}\lambda (k)$ where
$\lambda (k)$ is the
dispersion function which give the dependence of the spectral parameter
$\lambda$ on quasimomentum $k$ \cite{F}. Recall that the Bloch
function $\psi(x,\lambda)$ is a quasiperiodic one and
$\psi(x+a,\lambda)= e^{ikx} \psi(x,\lambda)$.

Thus the integrated density of states for the operator $H_0$ is
nondecreasing function which equals 0 at $\lambda <
\inf\sigma(H_0)=\lambda_0$ and is constant on each gap
$\overline{\Delta}_j=(\lambda_{2j-1}, \lambda_{2j})$. It is known that
$$
N(\lambda)=\frac{\tilde k(\lambda)}{\pi}
$$
where $\tilde k(\lambda)=\Re k(\lambda)$ and $k(\lambda)$ is the
global
quasimomentum introduced in \cite{F}. The quasimomentum $k(\lambda)$ is
real in bands and has nontrivial imaginary part inside gaps.

We introduce the integrated density of states for the operator $H$
given by (\ref{H}) as follows
\begin{equation}
N_{\varepsilon}(E) =
\frac{1}{a}\int\limits_{0}^{a}e_{\varepsilon}(x,x,E)dx
\label{DS}
\end{equation}
where $e_{\varepsilon}(x,y,E)$ is the corresponding spectral function. In
\cite{B1,B2} it has been shown that $e_{\varepsilon}(x,y,E)$ has the form
\begin{equation}
e_{\varepsilon}(x,y,E)=\frac{1}{2\pi}\int\limits_{-\infty}^{E}
\frac{g(x,E^{\prime})g(y,E^{\prime})}{|M(E)|^2}dE^{\prime}.
\label{K}
\end{equation}
Here $g(x,E)$ is a solution of equation $Hg=Eg$ which satisfies the
condition $g\to 0$ as $x\to -\infty$. Let us  normalize it by the
asymptotic relation~
\begin{equation}
g(x,E)\sim \frac{c}{\sqrt[4]{|E+\varepsilon x|}}
e^{-\frac{1}{\varepsilon}\frac{2}{3}|E+\varepsilon x|^{3/2}}
\label{g}
\end{equation}
where the real constant $c$ will be fixed later. The function $M(E)$ is
the so-called Jost function and is defined by the equation
\begin{equation}
M(E)=\frac{W[g, f]}{W[f, f]}.
\label{M}
\end{equation}
Here $W[g,f]\equiv gf'-g'f$ and $f(x,E)$ is the solution
of the same equation having the oscillating behaviour as $x\to +\infty$
and normalized by the asymptotic relation
$$
f(x,E)\sim \frac{1}{\sqrt{2}\sqrt[4]{E+\varepsilon x}}
e^{\frac{i}{\varepsilon}\frac{2}{3}(E+\varepsilon x)^{3/2}}
$$

From equations (\ref{K}) and (\ref{M}) it is clear that
$e_{\varepsilon}(x,y,\lambda)$ does not depend on the choice of the
constant $c$ in (\ref{g}). However it is convenient to fix the
normalization of the Jost function $M(E)$ unambiguously. Since we shall
study the behaviour of $N_{\varepsilon}(E)$ end hence of
$e_{\varepsilon}(x,x,E)$ as $\varepsilon \to 0$ it is enough to fix the
constant $c$ in the leading order of $\varepsilon$. Let us set
\begin{equation}
c=\exp{\left[ -\frac{1}{\varepsilon}\left(
\int\limits_{-\infty}^{\lambda_0}|k(\lambda)|d\lambda -
\int\limits_{-\infty}^{0}\sqrt{-\lambda}d\lambda \right)-
\int\limits_{-\infty}^{\lambda_0}\Phi(\lambda)d\lambda \right]}
(1+O(\varepsilon))
\label{c}
\end{equation}
where
\begin{equation}
\Phi(\lambda)=\frac{1}{a}\langle\partial_{\lambda}\phi(x,\lambda),
\phi(x,\lambda)\rangle,
\label{BF}
\end{equation}
$\phi(x,\lambda)$ is the periodic part of the Bloch function $\psi$
normalized by the condition (\ref{N}) and $\langle\cdot, \cdot\rangle$ is
the standard inner product in $L_2(0, a)$. Since $|k(\lambda)|\sim
\sqrt{|\lambda|}$ as $|\lambda|\to \infty$, the difference of integrals in
(\ref{c}) converges.

To study the behaviour of the integrated density of states for
Bloch electrons in electric field (\ref{DS}) at the limit of vanishing
strength $\varepsilon$ we have to study the asymptotics of the
spectral density  $|M(E)|^{-2}$ and of the solution $g(x,E)$ at
real values of $E$.

\section{An asymptotic behaviour of the spectral density.}

In \cite{B1,B2} there has been obtained the asymptotic formula
for $M(E)$ which allowed to study the structure of roots of
$M(E)$ in the lower half-plane (resonances) and to estimate
their imaginary parts. However the formulae
obtained there can not be effectively used to study the behaviour of the
spectral density $|M(E)|^{-2}$ as $\varepsilon\to 0$ on the
spectrum and hence the behaviour of the integrated density of
states (\ref{DS}). As it has been mentioned in Introduction in
the present paper we propose a
novel method of analysis of the Jost function. The advantage of
the asymptotic formulae obtained below is that although we study
the spectral density only in the leading order of $\varepsilon$
we artificially preserve a number of terms which are
exponentially small with respect to $\varepsilon$. However it is
these terms which allow us to effectively study the behaviour
of the integrated density of states as $\varepsilon \to 0$.

In the case of $N$-gap potential the Jost
function $M(E)$ is described in the leading order of
$\varepsilon$ as follows
\begin{equation}
M(E)=e^{i\xi}\overline{F_{N}}(E)[1+O(\varepsilon)].
\label{MAS1}
\end{equation}
The function $F_{N}(E)$ is given by the matrix product
\begin{equation}
\left( \begin{array}{c}
F_N\\
\overline{F_N}
\end{array}\right) =
X_N\ldots X_1
\left(\begin{array}{c}
F_0\\
\overline{F_0}
\end{array}\right),
\label{PROD}
\end{equation}
where $F_0 = -i$ and $\xi$ is some phase factor.

The matrices
\begin{equation}
X_l(E)=
\left(\begin{array}{cc}
\alpha_l&\beta_l\\
\bar{\beta_l}&\bar{\alpha_l}
\end{array}\right),
\label{X}
\end{equation}
$l=1,\ldots , N$, have the entries
\begin{equation}
\alpha_l = -iP_l^{-1/2}e^{i(\omega-\omega_l)},
\label{ENTR1}
\end{equation}
\begin{equation}
\beta_l = -i\sqrt{P_l^{-1}-1}e^{-i(\omega-\omega_l)}.
\label{ENTR2}
\end{equation}
Here
\begin{equation}
\omega=\frac{E\pi}{\varepsilon a}
\label{W}
\end{equation}
is the rescaled energy and $\omega_l$ is the constant:
\begin{equation}
\omega_l = \int\limits_{\pi (l-1)/a}^{\pi l/a}
(\frac{1}{\varepsilon}\lambda(k)+\Phi(k))dk,
\label{WL}
\end{equation}
where $\Phi(k)dk\equiv\Phi(\lambda)d\lambda$ with $\Phi(\lambda)$ given by
(\ref{BF}). The integration in (\ref{WL}) is performed over the $l$-th
Brillouin zone $\left[ \frac{\pi (l-1)}{a}, \frac{\pi l}{a}\right]$ which
corresponds to the $l$-th spectral band $\Delta_l$ under the mapping
$k(\lambda)$.

The constant $P_l$ is given as follows
\begin{equation}
P_l=e^{-2\pi\mu_l}
\label{PROB}
\end{equation}
with
\begin{equation}
\mu_l = \frac{1}{\varepsilon}\int\limits_{\bar{\Delta_l}}|\Im
k(\lambda)|d\lambda.
\label{MU}
\end{equation}
Here $\overline {\Delta_l}=(\lambda_{2l-1}, \lambda_{2l})$ is the $l$-th gap in
the spectrum of $H_0$. It should be noted that $P_l$ is the probability of
transmission of an electron under the influence of applied electric field
from the conducting band $\Delta_l$ to the band $\Delta_{l+1}$
( see \cite{B3, B4, G1} and references therein).

The phase factor $\xi$ in (\ref{MAS1}) also can be described explicitly.
However we do not need its knowledge since to study the integrated density
of states $N_{\varepsilon}(E)$ (\ref{DS}) we need $|M(E)|$ only.

In what follows we show how to simplify the matrix product (\ref{PROD}).
Let us introduce the parameter
\begin{equation}
\sigma_l = \frac{1}{4}\ln (1-P_l)^{-1}
\label{SIG}
\end{equation}
On the basis of (\ref{PROB}) and (\ref{MU}) one sees that
\begin{equation}
\sigma_i\sim \frac{1}{4}e^{-2\pi\mu_l}\to 0\qquad as\quad \varepsilon\to 0
\label{SIGAS}
\end{equation}
Thus the quantity $\sigma_l$ can be considered as an additional parameter
of the problem which is exponentially small in $\varepsilon$. Let
$$
\sigma = \max_{l}\lbrace\sigma_l\rbrace.
$$
Then the following statement is valid.

{\bf Proposition 1.} {\it Let the function $F_l(E)$ be defined
by the matrix product (\ref{PROD}). Then in the leading order
of $\sigma$
\begin{equation}
\frac{F_l(E)}{\overline{F_l}(E)} = -1 + O(\sigma)
\label{RAT}
\end{equation}
This estimate is uniform w.~r.~t. $E\in \RR$.
}

\underline{Proof.} This statement can be proved by the mathematical
induction method.  For $l=1$ one has $$
\frac{F_1(E)}{\overline{F_1}(E)}=\frac{e^{-2\sigma-i(\omega-\omega_1)}-
e^{i(\omega-\omega_1)}}{e^{i(\omega-\omega_1)-2\sigma}-
e^{-i(\omega-\omega_1)}}
=\frac{1-e^{2i(\omega-\omega_1)}+O(\sigma)}{e^{2i(\omega-\omega_1)}+O(\sigma)}=
-1+O(\sigma).
$$
Here we used the fact that $F_0=-i$ and in the second equality we expanded
$e^{-2\sigma}$ into the Taylor series as $\sigma\to 0$.

Now suppose that the statement is valid for $l-1$.
Since $(F_l, \overline{F_l})_T = X_l(F_{l-1}, \overline{F_l})^T$ one has that
\begin{equation}
\overline{F_l}=i \left( 1-e^{-4\sigma_l}\right)^{-1/2}
\left[ e^{i(\omega-\omega_1)-2\sigma_l}F_{l-1}+
e^{-i(\omega-\omega_1)}\overline{F_{l-1}} \right].
\label{A1}
\end{equation}
Hence
\begin{equation}
\frac{F_l}{\overline{F_l}} = -\frac{F_l}{\overline{F_l}}
\left(
\frac{e^{-2\sigma_l}\frac{\overline{F_{l-1}}}{F_{l-1}}+
e^{2i(\omega-\omega_1)}}
{e^{2i(\omega-\omega_1)}\frac{F_{l-1}}{\overline{F_{l-1}}}+1}
\right)
\label{RAT1}
\end{equation}
Now expanding $e^{-2\sigma_l}$ as $\sigma_l\to 0$ and using the
induction supposition we obtain (\ref{RAT}).$\Box$

On use of this proposition  one can obtain the asymptotics of
$\overline{F_l}$ as $\sigma \to 0$.

{\bf Lemma 1.}
{\it
On the real  axis $E$ the functions $\overline{F_l}, l=1,\ldots , N$, in
the leading order of $\sigma$ are given as follows
\begin{equation}
\overline{F_l(E)}=e^{i\tilde {\phi}_l(E)}
\prod_{j=1}^{l}\left( 1-e^{-4\sigma_j}\right)^{-1/2}
\left( e^{2i(\omega-\omega_j)-2\sigma_j}-1\right) [1+O(\sigma)],
\label{F}
\end{equation}
where
\begin{equation}
\tilde {\phi}_l(E) = -\frac{\pi l}{2}\, -\,  \sum_{j=1}^{l}(\omega-\omega_1).
\label{A2}
\end{equation}
}

\underline{Proof.} Take out from square brackets in (\ref{A1}) the
multiplier $\overline{F_{l-1}}$, and then use the recursion
procedure. This yields:
$$ \overline{F_l}=(i)^l
\prod_{j=1}^{l}\left( 1- e^{-4\sigma_i}\right)^{-1/2} \prod_{j=1}^{l}\left[
e^{i(\omega-\omega_j)-2\sigma_j}
\left(\frac{F_{j-1}}{\overline{F_{j-1}}}\right) +
e^{-i(\omega-\omega_j)}\right].
$$
Now the statement of the lemma is obtained easily on use of the asymptotic
relation (\ref{RAT}) for the ratio $(F_{j-1}/\overline{F_{j-1}})$.
$\Box$

Taking now the absolute value of the r.h.s. of (\ref{F}) one
easily obtains $|F_N|$. To write down it in a compact form let
us introduce the function
\begin{equation}
\delta (\omega,\omega
',\sigma)=\frac{\tanh{\sigma}}{\sin^2{(\omega-\omega
')}+\tanh^2{\sigma}\cos^2{(\omega-\omega ')}}
\label{DELTA}
\end{equation}
and the functions
\begin{equation}
\delta_j(E)\equiv\delta(\frac{E\pi}{\varepsilon a}, \omega_j, \sigma_j),
\quad j=1, \ldots , N
\label{DELTAj}
\end{equation}
(recall that $\omega_j$ and $\sigma_j$ are given by (\ref{WL})
and (\ref{SIG}) respectively). Then
\begin{equation}
\frac{1}{|F_N(E)|^2}= \prod_{l=0}^{N}\delta_l(E)(1+O(\sigma)).
\label{FN}
\end{equation}
Now on use of the asymptotics (\ref{MAS1}) one obtains the
main statement of this section.

{\bf Theorem 1. }
{\it In the leading order of $\varepsilon$ the spectral density
$|M(E)|^{-2}$ is given as follows
\begin{equation}
\frac{1}{|M(E)|^2}=\prod_{j=1}^{N}\delta_j(E)(1+O(\varepsilon)),
\label{MAS2}
\end{equation}
where the functions $\delta_j$ are defined by (\ref{DELTAj}).
}

\section{Weak limit of the spectral density.}

The asymptotic formula (\ref{MAS2}) obtained in the previous
section allows to prove that the spectral density $|M(E)|^{-2}$ in a
weak sense as $\varepsilon \to 0$ tends to unit. This important
fact will be essentially used in the study of the integrated
density of states (\ref{DS}).

{\bf Theorem 2.}
{\it Let $\Delta E$ be an interval of the real axis $E$ and $f(E)$ be an
arbitrary smooth function defined on it. Then}
\begin{equation}
\lim_{\varepsilon\to 0}\int\limits_{\Delta E}
\frac{f(E)dE}{|M(E)|^2} =
\int\limits_{\Delta E}f(E)dE.
\label{WLM}
\end{equation}
On use of the asymptotic relation (\ref{MAS2}) it is enough to show that
\begin{equation}
\lim_{\varepsilon\to 0}\int\limits_{\Delta E}
\left( \prod_{j=1}^{N}\delta_j(E)\right) f(E)=
\int\limits_{\Delta E}f(E)dE.
\label{WLPRD}
\end{equation}

Will shall prove the latter fact by several steps.

{\bf Proposition 2.} {\it Let $\Delta\omega_n = [\omega '+\pi n-\pi/2,
\omega '+\pi n -\pi/2]$, $\omega '$ be the fixed point, $n$ be an arbitrary
integer and $f(\omega)$ be a smooth function on $\Delta\omega_n$. Let
$\delta(\omega, \omega ', \sigma)$ be defined by (\ref{DELTA}). Then}
\begin{equation}
\lim_{\sigma\to 0}\frac{1}{\pi}\int\limits_{\Delta\omega_n}
\delta(\omega, \omega ', \sigma)f(\omega)d\omega = f(\omega '+\pi n)
\label{A2}
\end{equation}

\underline{Proof.} Make the change of variables
$$
z=\tg{(\omega -\omega '- \pi n)}
$$
and denote
$$
\hat {\sigma} =\tanh{\sigma}, \qquad \hat f(z)=f(z+\omega '+\pi n)
$$
Then
$$
\lim_{\sigma\to 0}\frac{1}{\pi}\int\limits_{\Delta\omega_n}
\delta(\omega, \omega ', \sigma)f(\omega)d\omega
 = \lim_{\sigma\to 0}\frac{1}{\pi}
\int\limits_{-\infty}^{\infty}
\frac{\hat {\sigma}\hat f(z)}{\hat {\sigma}^2+z^2}dz.
$$
Due to the fact that $\hat {\sigma}/(\hat {\sigma}^2+z^2)$ as $\sigma\to 0$
is the $\delta$-type sequence  the $r.h.s.$ of the latter
equals $\hat f(0)= f(\omega '+\pi n)$. $\Box$

{\bf Lemma 2.} {\it Let $\Delta E$ be an real interval, $f(E)$ be a
smooth function and $\delta_j(E)$ is given by (\ref{DELTAj}). Then}
\begin{equation}
\lim_{\varepsilon\to 0}\int\limits_{\Delta E}\delta_j(E)f(E)dE =
\int\limits_{\Delta E}f(E)dE.
\label{WLD}
\end{equation}

\underline{Proof.}  On use of the definition of $\delta_j$ one has
\begin{equation}
\int\limits_{\Delta E}\delta_j(E)f(E)dE =
\frac{\varepsilon a}{\pi}\int\limits_{\Delta\omega}
\delta(\omega, \omega_j, \sigma_j)\tilde f(\omega)d\omega,
\label{A3}
\end{equation}
where
$$
\Delta\omega=\frac{\pi\Delta E}{\varepsilon a}\quad and \quad
\tilde f(\omega)=f(\frac{\varepsilon a\omega}{\pi}) .
$$

Let us split the interval $\Delta E$ on subintervals which have the length
$\varepsilon a$ and the middle points
$$
E_j^{(n)}=\frac{\varepsilon a}{\pi}\omega_j+\varepsilon an\equiv
\frac{a}{\pi}\int\limits_{\Delta_j}[\lambda(k)+\varepsilon \Phi(k)]dk+
\varepsilon an
$$
(here $n\in \ZZ$ is such that $E_j^{(n)}\in \Delta E$). Under
this procedure the
interval $\Delta\omega = \frac{\varepsilon a}{\pi}\Delta E$ is splitted
into subintervals $\Delta\omega_n^{(j)}=[\omega_j+\pi n-\pi /2,
\omega_j+\pi n-\pi /2]$ where $\omega_j$ is given by (\ref{WL}).
Thus the integral in the r.~h.~s. of (\ref{A3}) can
be presented as the sum of integrals over $\Delta\omega_n^{(j)}$.
According to proposition 2 each of them as $\sigma_j\to 0$ tends to
$\pi \tilde f(\omega_j+\pi n)\equiv\pi f(E_j+\varepsilon an)$. Thus the
r.~h.~s. of (\ref{A3}) takes the form
$\varepsilon a\sum_{n}'f(E_j+\varepsilon an)$ where summation is performed
over such $n$ that $E_j^{(n)}=E_j+\varepsilon an\in \Delta E$. Obviously
as $\varepsilon\to 0$ this sum gives the integral in the r.~h.~s. of (\ref{WLD}).
$\Box$

The statement of lemma 2 differs from that we need to prove,
namely equation (\ref{WLPRD}), by the change of the single multiplier
$\delta_j(E)$ by their product. In this connection we have to prove that
in appropriate sense for some $j$
\begin{equation}
\prod_{l=0}^{N}\delta_l(E)\sim\delta_j(E)\qquad as\quad \varepsilon\to 0
\label{APR}
\end{equation}

Due to the singular behaviour of $\delta_j(E)$ as $\varepsilon\to 0$ the
direct analytic proof of (\ref{APR}) meets serious difficulties.
Therefore we shall use the relation (\ref{FN}),
take into account that the function $F_N(E)$ is defined by the
matrix product (\ref{PROD}) and
perform more detailed analysis of the latter.

Together with functions $F_N(E)$ let consider the functions $F_N^{(l)}(E)$
which we define as follows
$$
\left(\begin{array}{cc}
F_N&F_N^{(l)}\\
\overline{F_N}&\overline{F_N^{(l)}}
\end{array}\right)=
X_N\ldots X_l
\left(\begin{array}{cc}
F_{l-1}&\frac{i}{2}F_{l-1}\\
\overline{F_{l-1}}&-\frac{i}{2}\overline{F_{l-1}}
\end{array}\right)
$$
On use of (\ref{ENTR1}),(\ref{ENTR2}) one has that
$$
\det{X_l = 1}.
$$
Hence
$$
F_{N}\overline{F_N^{(l)}}-\overline{F_N}F_{N}^{(l)}=-i|F_{l-1}|^2
$$
and therefore
\begin{equation}
\left|\frac{F_{l-1}}{F_N}\right|^2=
-2\Im\left(\frac{\overline{F_N^{(l)}}}{\overline{F_N}}\right).
\label{A4}
\end{equation}

Now one can easily obtain the asymptotic formula for
$\overline{F_N^{(l)}}(E)$. To this end repeat the proof of
lemma 1 taking into account the additional multiplier $-\frac{i}{2}$
in definition of $\overline{F_N^{(l)}}(E)$. The result is the
following statement.

{\bf Proposition 3.} {\it In the leading order of $\sigma$ the functions
$\overline{F_N^{(l)}}(E)$ have the form}
$$
\overline{F_N^{(l)}}(E)=ie^{i\tilde {\phi}_N(E)}\prod_{j=1}^{N}
\left(1-e^{-4\sigma_j}\right)^{-1/2}
\left( e^{2i(\omega-\omega_l)-2\sigma_l}+1\right)*
$$
$$
\prod_{{j=1}\atop{j\ne l}}^{N}
\left(e^{2i(\omega-\omega_j) -2\sigma_l}-1\right)
[1+O(\sigma)].
$$

Compare now the obtained formula for $\overline{F_N^{(l)}}$ with the
equation (\ref{F}) for $\overline{F_N}$. Obviously
$$
\frac{\overline{F_N^{(l)}}}{\overline{F_N}}=
\frac{i\left( e^{2i(\omega-\omega_l)-2\sigma_l}+1\right)}
{\left( e^{2i(\omega-\omega_l)-2\sigma_l}-1\right)}
[1+O(\sigma)]
$$
Inserting this representation into (\ref{A4}) and performing simple
calculations one obtains the following statement.

{\bf Lemma 3.} {\it  On the real axis $E$ in the leading order of $\sigma$
\begin{equation}
\left|\frac{F_{l-1}(E)}{F_N(E)}\right|^2=\delta_l(E)(1+O(\sigma))
\label{COMP}
\end{equation}
when $\delta_l(E)$ is given by (\ref{DELTAj}).}

The estimate (\ref{APR}) is just the simple consequence of
(\ref{COMP}). Indeed,
set $l=1$ and note that $F_0=-i$. Then
\begin{equation}
\prod_{j=1}^{N}\delta_j(E)[1+O(\sigma)]=
\left|\frac{1}{F_N(E)}\right|^2=
\delta_1(E)[1+O(\sigma)]
\label{COMP1}
\end{equation}
Here the first equality is due to (\ref{FN}) and the second follows from
(\ref{COMP}).

Now the statement of the theorem formulated in the beginning of
the present section is the direct consequence of lemma 2 and the
relation (\ref{COMP1}).

\section{Nonuniform asymptotics of the solution $g(x, E)$}

In this short section we
essentially use the results of the paper \cite{B1}. However we need this section since
the nonuniform w.~r.~t. $E$ asymptotics of solution $g(x,E)$ combined with
the behaviour of $M(E)$ studied in the previous section define the
asymptotic behaviour of the integrated density of states (\ref{DS}).

Quasiclassical analysis of equation $Hf=Ef$ performed in
\cite{B1, B2, B3, B4, B5} leads to
an important link between the dispersion function $\lambda(k)$ of the
operator $H_0$ and the variable $E+\varepsilon x$ which naturally arises
in the asymptotics of $g(x,E)$ given by (\ref{g}) as $x\to
-\infty$. This the link has
form
\begin{equation}
\lambda(k)=E+\varepsilon x
\label{IDENT}
\end{equation}
and is valid for all $x$. At any fixed $x$ this link can be considered as
identification between the real axis $E$ and the spectral axis $\lambda$
of the operator $H_0$. Let us denote by
\begin{equation}
\Delta_l(\varepsilon)=
[\lambda_{2l-2}-\varepsilon x, \lambda_{2l-1}-\varepsilon x]
\label{BAND}
\end{equation}
the shifted $l$-th band of $H_0$
By
\begin{equation}
\overline{\Delta_l(\varepsilon)}=
(\lambda_{2l-2}-\varepsilon x, \lambda_{2l-1}-\varepsilon x)
\label{GAP}
\end{equation}
we denote the shifted $l$-th gap.

As $\varepsilon\to 0$ the asymptotics of $g(x,E)$ turns out to be
essentially different at $E\in \Delta_l(\varepsilon)$ and at
$E\in\overline{\Delta_l(\varepsilon)}$. Inside the shifted bands
$\Delta_l(\varepsilon)$
\begin{equation}
g(x,E)=(A_l(E)f_l(x,E)+\bar A_l\overline{f_l(x,E)})[1+O(\varepsilon)].
\label{gB}
\end{equation}
Here
\begin{equation}
A_l(E)=e^{i\eta}F_{l-1}(E)
\label{Al}
\end{equation}
and
\begin{equation}
f_l(x,E) = (\dot {\lambda}(k))^{-1/2}\exp\left[
\int\limits_{\lambda_{2l-2}}^{E+\varepsilon x}\left(
\frac{i}{\varepsilon}k(\lambda)+\Phi(\lambda)\right)d\lambda\right]
\phi(x,\lambda)
\label{fl1}
\end{equation}
The phase factor $\eta$ in (\ref{Al})can be written down explicitly but we
do not need it below. The function $\phi(x,\lambda)$ in (\ref{fl1}) is
a periodic part of the Bloch solution $\psi(x,\lambda)$ normalized by the
condition (\ref{N}). The variables $k$ and $\lambda$ depend on
$(E+\varepsilon x)$ according to the relation (\ref{IDENT})  and
$\Phi(\lambda)$ is given by (\ref{BF}).

\underline{Remark.} The asymptotics (\ref{gB}) is valid outside
the $\varepsilon^{1/3}$-vicinity of the shifted edges $\lambda_l -
\varepsilon x$ \cite{B1, B3}. In the vicinity of these points
one has to use the local asymptotics \cite{B5}.

In the shifted gaps $\overline{\Delta_l(\varepsilon)}$ outside the
$\varepsilon^{1/3}$-vicinity of the shifted edges $\lambda_l -
\varepsilon x$ the solution $g$ has the form
\begin{equation}
g(x,E)=(H_l(E)h_l^-(x,E)+G_l(E)h_l^+(x,E))[1+O(\varepsilon)]
\label{gG}
\end{equation}
The coefficients $H_l(E)$ and $G_l(E)$ are related with coefficients
$A_l$, $\bar A_l$ as follows
\begin{equation}
H_l=\frac{1}{2}\left[ e^{i\pi /4}e^{i(\omega-\omega_l)}A_l+
e^{-i\pi /4}e^{-i(\omega-\omega_l)}\bar A_l\right]
\label{Hl}
\end{equation}
\begin{equation}
G_l=\left[ e^{-i\pi /4}e^{i(\omega-\omega_l)} A_l+
e^{i\pi /4}e^{-i(\omega-\omega_l)}\bar A_l\right]
\label{Gl}
\end{equation}
Recall that $\omega$ and $\omega_l$ are given by (\ref{W}) and (\ref{WL})
respectively.

The functions $h_{l}^{\pm}(x,E)$ have the form
\begin{equation}
h_{l}^{\pm}(x,E)=|\dot {\lambda}(k)|^{-1/2}\exp{\left[
\pm\int\limits_{\lambda_{2l-1}}^{E+\varepsilon x}(\frac{1}{\varepsilon}
|\Im k(\lambda)|+\Phi(\lambda)d\lambda\right]}\phi_{\pm}(x,\lambda))
\label{hl}
\end{equation}
The function $h^+$ exponentially increases and $h^-$ exponentially
decreases as $\varepsilon\to 0$ inside the interval
$\overline{\Delta_l(\varepsilon)}$.

\section{The asymptotics of spectral projections kernels.}

Let us denote by $e_{\varepsilon}(x,y,\Delta E)$ the kernel of the
spectral projection of the operator $H$ on the interval $\Delta E$. Due to
(\ref{K}) it has the form
\begin{equation}
e_{\varepsilon}(x,y,\Delta E)=\frac{1}{2\pi}\int\limits_{\Delta E}
\frac{g(x,E)g(y,E)}{|M(E)|^2}
\label{SP}
\end{equation}

\noindent
{\bf Theorem 3.} {\it Let
$\Delta_l^{\prime}(\varepsilon)\subset\Delta_l(\varepsilon)$ be a
subinterval of the shifted $l$-th band $\Delta_l(\varepsilon)$
 such that it does not contain
the $\varepsilon^{2/3}$-vicinity of the edges and
 $\overline{\Delta_l^{\prime}(\varepsilon)} \to
\overline{\Delta_l}$ as $\varepsilon
\to 0$ where $\overline{\Delta_l}$ is the $l$-th band of $H_0$.
 Then at bounded $|x|,|y|$
$$ \lim_{\varepsilon\to
0}e_{\varepsilon}(x,y,\Delta_l^{\prime}(\varepsilon))=e_l(x,y)
$$
where $e_l(x,y)$ is the kernel of the spectral projection of the operator
$H_0$ on the $l$-th band $\Delta_l\equiv\Delta_l(0)$.
}

\underline{Proof. } Let us rewrite (\ref{fl1}) in the form
\begin{equation}
f_l(x,E)=(\dot {\lambda}(k))^{-1/2}\psi(x,\lambda)e^{iS(x,E,\varepsilon)}
\label{fl2}
\end{equation}
where
$$
S(x,E,\varepsilon)=\frac{1}{\varepsilon}(E+\lambda_{2l-2})k-
\int\limits_{\frac{\pi(l-1)}{a}}^{K}\left(\frac{1}{\varepsilon}\lambda(k)+
\Im\Phi(k)\right)dk.
$$
Since $k$ depends on $(E+\varepsilon x)$ accordingly to
(\ref{IDENT}) then on the basis of (\ref{gB}), (\ref{fl2}) one has that at
$E\in\Delta_l(\varepsilon)$
\begin{equation}
g(x,E)g(y,E)=(T_1+T_2)(1+O(\varepsilon))
\label{gg}
\end{equation}
where
\begin{equation}
T_1=\frac{2\Re[\psi(x,\lambda)\overline{\psi(y,\lambda)}]}
{\dot {\lambda}(k)}|A_l(E)|^2
\end{equation}
\label{T1}
\begin{equation}
T_2=2\dot {\lambda}^{-1}\Re\left[\psi(x,\lambda)\psi(y,\lambda)
e^{2iS(x,E,\varepsilon)}A_l^2(E)\right].
\label{T2}
\end{equation}

Consider the contribution of each term in (\ref{gg}) into the kernel
$e_{\varepsilon}(x,y,\Delta_l(\varepsilon))$ given by (\ref{K}). Taking
into account that $|A_l|=|F_{l-1}|$, $|M|=|F_{N}|$ and using the statement
of Lemma 3, namely the estimate (\ref{COMP}), one obtains
\begin{equation}
\frac{1}{ 2\pi}\int\limits_{\Delta_l^{\prime}(\varepsilon)}\frac{T_1(x,E)}
{|M(E)|^2}dE=\frac{1}{\pi}\int\limits_{\Delta_l^{\prime}(\varepsilon)}
\frac{\Re[\psi(x,\lambda)\overline{\psi(y,\lambda)}]}{\dot {\lambda}(k)}
\delta_l(E)[1+O(\varepsilon)dE
\label{A5}
\end{equation}

Make use of (\ref{WLD}) and note that $\Delta_l^{\prime}(\varepsilon)\to
\Delta_l$ and $E\to\lambda$ as $\varepsilon\to 0$. Then it is obvious that
the limit of r.~h.~s. of (\ref{A5}) is exactly the kernel of spectral
projection $e_l(x,y)$ of $H_0$ on the $l$-th band given by
(\ref{KH0}).  Consequently the
statement of the theorem is valid if
$$
\frac{1}{2\pi}\int\limits_{\Delta_l^{\prime}(\varepsilon)}
\frac{T_2(x,E)}{|M(E)|^2}dE \longrightarrow 0\quad as\quad \varepsilon\to
0.
$$
However this fact is quite obvious due to the rapid oscillations of the
phase factor $\exp{[iS(x,E,\varepsilon)]}$ as $\varepsilon\to 0$.
$\Box$

{\bf Theorem 4.}{\it
Let
$\overline{\Delta_l^{\prime}(\varepsilon)}
\subset\overline{\Delta_l(\varepsilon)}$ be a subinterval of the
shifted $l$-th gap $\overline{\Delta_l(\varepsilon)}$ given by
(\ref{GAP}) such that it does not contain
the $\varepsilon^{2/3}$-vicinity of the edges and
 $\overline{\Delta_l^{\prime}(\varepsilon)} \to
\overline{\Delta_l}$ as $\varepsilon
\to 0$ where $\overline{\Delta_l}$ is the $l$-th gap of $H_0$.
 Then at bounded $|x|,|y|$
$$
\lim_{\varepsilon\to
0}e_{\varepsilon}(x,y,\overline{\Delta_l^{\prime}(\varepsilon)})\,=\,0.
$$
}

\underline{Proof.}  At $E\in\overline{\Delta_l(\varepsilon)}$ the
solution $g(x,E)$ is given by (\ref{gG}). For the sake of simplicity
let us assume that $x=y$ (note that in studying the integrated
density of states we need namely this case). Then
\begin{equation}
g^2(x,E)=(G_l^2(h_l^+)^2+2H_lG_lh_l^+h_l^-+H_l^2(h_l^-)^2)[1+O(\varepsilon)]
\label{A6}
\end{equation}
Let us subsequently consider the contribution to
$e_{\varepsilon}(x,x,\overline{\Delta_l^{\prime}})$ of each term in
(\ref{A6}). On use of (\ref{SP}) and (\ref{A6}) the contribution of the
first term can be written as
$$
\int\limits_{\overline{\Delta_l(\varepsilon)}}\frac{G_l^2(E)h_l^+(x,E)^2}
{|M(E)|^2}=(-i(I_1-\bar I_1)+I_2)[1+O(\varepsilon)]
$$
where
$$
I_1=\int\limits_{\bar {\Delta}_l(\varepsilon)}e^{2i(\omega-\omega_l)}
e^{2\gamma(x,E,\varepsilon)}\frac{F_{l-1}^2(E)}{|F_N(E)|^2}q(x,E,\varepsilon)dE
$$
and
$$
I_2=\int\limits_{\bar
{\Delta}_2(\varepsilon)}e^{2\gamma(x,E,\varepsilon)}
\frac{|F_{l-1}(E)^2|}{|F_N(E)|^2}q(x,E,\varepsilon)dE
$$
Here
$$
\gamma(x,E,\varepsilon)=\frac{1}{\varepsilon}\int\limits_{\lambda_{2l-1}}^{E+\varepsilon x}
|\Im k(\lambda)|d\lambda >0
$$
and
$$
q(x,E,\varepsilon)=\frac{\phi^2(x,k)}{|\dot {\lambda}(k)|}\exp{
\left[2\int\limits_{\lambda_{2l-1}}^{E+\varepsilon
x}\Phi(\lambda)d\lambda\right]} .
$$
Now make simple estimates of $I_2$.
Recall that
\begin{equation}
\left|\frac{F_{l-1}(E)}{F_N(E)}\right|^2 =-\delta_l(E)(1+O(\sigma)).
\label{A7}
\end{equation}
Then notice that at $E\in\overline{\Delta_l(\varepsilon)}$:
$$
\gamma(x,E,\varepsilon) \le \frac{1}{\varepsilon}
\int\limits_{\lambda_{2l-1}}^{\lambda_{2l}}|\Im
k(\lambda)|d\lambda\equiv\pi\mu_l
$$
end hence
\begin{equation}
e^{[2\gamma(x, E,\varepsilon)]}\le
e^{2\pi\mu_l}\equiv\frac{1}{1-e^{-4\sigma_l}}
\label{A8}
\end{equation}
Inserting (\ref{A7}), (\ref{A8}) into the expression for $I_2$ one obtains
$$
|I_2|\le Q\int\limits_{\bar
{\Delta}_l^{\prime}(\varepsilon)}\frac{\delta_l(E)}{\left(
1-e^{-4\sigma_l}\right)}dE=
$$
$$
Q\int\limits_{\bar {\Delta}_l^{\prime}(\varepsilon)}
\frac{\tanh \sigma_l}{\left(1-e^{-4\sigma_l}\right)}
\frac{1}{\sin^2(\omega-\omega_l)+\tanh^2\sigma_l\cos(\omega-\omega_l)}dE,
$$
where $Q \ge |q(x,E,\varepsilon)|$.
Expanding the latter expression in powers of $\sigma_l$ as
$\sigma_l\to 0$ one gets
$$
I_2\le\frac{Q}{4}\int\limits_{\bar
{\Delta}_l^{\prime}(\varepsilon)}\frac{dE}{\sin^2(\omega-\omega_l)}
(1+O(\sigma_l))=
$$
$$
\frac{\varepsilon
aQ}{4\pi}\left[\ctg\left(\frac{\lambda_{2l}\pi}{\varepsilon
a}-\omega_l\right) -\ctg\left(\frac{\lambda_{2l-1}\pi}{\varepsilon
a}-\omega_l\right)\right](1+O(\varepsilon))
$$
Thus
$$
I_2\to 0\qquad as \qquad \varepsilon\to 0.
$$
The integral $I_1$ also vanishes as $\varepsilon\to 0$. Indeed,
due the chain of estimates
$$
\frac{F_l^2}{|M|^2}=\left|\frac{F_l}{F_N}\right|^2\frac{F_l}{\overline{F_l}}
(1+O(\varepsilon))=-\left|\frac{F_l}{F_N}\right|^2(1+O(\varepsilon))=\\
-\delta_l(E)(1+O(\varepsilon))
$$
it differs from the integral $I_2$ by the presence of the
rapidly oscillating factor.
Thus we have shown that the contribution of the first term in
(\ref{A6}) into the kernel $e_{\varepsilon}(x,x,E)$ vanishes as
$\varepsilon\to 0$.

Consider briefly the contribution of the second term in
(\ref{A6}). Notice that
$$
H_lG_l=\frac{1}{2}\left(e^{2i(\omega-\omega_l)}A_l^2+e^{-2i(\omega-\omega_l)}
\bar A_l^2\right)
$$
and hence
$$
\int\limits_{\bar
{\Delta}_l(\varepsilon)}\frac{H_lG_lh_l^+h_l^-}{|M(E)|^2dE}
\,\,\, \to \,\,\,0
$$
because each term in this integral contains the rapidly oscillating factor which causes the
vanishing of the limits as $\varepsilon\to 0$.

Finally, the contribution of the third term in (\ref{A6}) into spectral
projections on $\overline{\Delta_l(\varepsilon)}$ also tends to 0. It
becomes quite clear if one compare this term with the first one.
Instead of exponentially growing factor
$e^{2\gamma(x,E,\varepsilon)}$ the third term exponentially decreases
as $e^{-2\gamma(x,E,\varepsilon)}$. Since we have proved that
contribution of the first term vanishes the vanishing of the third one
becomes obvious.

\section {Density of States and Integrated Density of States}

Along with the integrated density of states (\ref{DS}) one can consider
the density if states
$$
n_{\varepsilon}(E) = \frac{dN_{\varepsilon}(E)}{dE}.
$$
In our case it has the form
$$
n_{\varepsilon}(E)=\frac{1}{a}\int\limits_0^a \beta (x,E)dx,
$$
where
$$
\beta (x,E) = \frac{1}{2\pi}\frac{g^2(x,E)}{|M(E)|^2}.
$$
The density of states has no regular limit as $\varepsilon \to 0$.
Indeed as it has been shown in Sec.4 the spectral density $|M(E)|^{-2}$
is a sort of periodic $\delta$-type sequence and its limit is defined
only in a weak sense (\ref{WLM}). The product $g^2(x,E)$ due to (\ref{gg}),
 (\ref{T2}) also has no regular limit.

However the limit of the integrated density of states (\ref{DS}) as
$\varepsilon \to 0$ is well defined. This is the direct consequence of
the results of the previous section. Indeed if $E \in
\Delta_{m+1}^{\prime} (\varepsilon)$ (for the notations see
theorem 3) the integral (\ref{K}) at $y=x$ can be presented in the form
$$
e_{\varepsilon}(x,x,E) = \sum_{j=1}^m
\int\limits_{\Delta_{j}^{\prime}  (\varepsilon)} \beta(x,E)dE +
\sum_{j=0}^m
\int\limits_{\overline {\Delta_{j}^{\prime} (\varepsilon)}}
\beta(x,E)dE +
$$
\begin{equation}
\int\limits_{E_{2m}^{\prime}}^E \beta(x,E) dE + \sum_{j=0}^{2m} \int\limits_
{V_j^{\varepsilon}} \beta(x,E) dE.
\label{B1}
\end{equation}
Here $\Delta_{j}^{\prime} (\varepsilon)$
and $\overline {\Delta_{j}^{\prime} (\varepsilon)}$ are
subintervals of shifted bands and gaps described in theorems 3
and 4 respectively. By $V_j^{\varepsilon}$ we denote the
$\varepsilon^{2/3}$-vicinity of the points $\lambda_j
-\varepsilon x$
and $E_{2m}^{\prime} > \lambda_{2m}
-\varepsilon x \in V_{2m}^{\varepsilon}$.

As it follows from theorem 3 at the zero bias limit
\begin{equation}
\sum_{j=1}^m
\int\limits_{\Delta_{j}^{\prime} (\varepsilon)} \beta(x,E)dE
\,\,\, \to \,\,\, \sum_{j=1}^{m}e_j(x,x),
\label{B2}
\end{equation}
where $e_j(x,x)$ is given by (\ref{KH0}) and
\begin{equation}
\int\limits_{E_{2m}}^{E}\beta (x,E) dE \,\,\, \to
\int\limits_{\lambda_{2m}}^{\lambda}\frac{\Re [\psi(x,\lambda)
\overline{\psi(y,\lambda)}]}{\dot {\lambda}(k)}d\lambda.
\label{B3}
\end{equation}
The consequences of
theorem 4 is the limit
\begin{equation}
\sum_{j=0}^m
\int\limits_{\overline  {\Delta_{j}^{\prime} (\varepsilon)}}
\beta(x,E)dE \to 0.
\label{B4}
\end{equation}

One also can prove that
\begin{equation}
\int\limits_
{V_j^{\varepsilon}} \beta(x,E) dE \,\,\, \to \,\,\,0.
\label{B5}
\end{equation}
To this end one needs a more detailed analysis of the behaviour
of $\beta(x,E)$ in the $\varepsilon^{2/3}$-vicinity of the
shifted edges $\lambda_j -\varepsilon x$. This can be done
explicitly, however it is out the scope of the present paper.

Finally inserting the limits (\ref{B2}) (\ref{B3}) (\ref{B4}) (\ref{B5})
into (\ref{B1}) one obtains that at zero bias limit
\begin{equation}
e_{\varepsilon}(x,x,E)\,\,\, \to \,\,\, \sum_{j=1}^{m}e_j(x,x) +
\frac{1}{\pi}\int\limits_{\lambda_{2m}}^{\lambda}
\frac{\Re [\psi(x,\lambda)\overline{\psi(x,\lambda)}]}{\dot
 {\lambda}(k)}d\lambda \equiv e(x,x,\lambda).
\label{LB}
\end{equation}
Here $E$ belongs to the $m+1$-th
shifted band.

Analogously one can prove that if $E$ belongs to the shifted
$m+1$-th gap at the zero bias limit
\begin{equation}
e_{\varepsilon}(x,x,E)\,\,\, \to \,\,\, \sum_{j=1}^{m}e_j(x,x)
\label{LG}
\end{equation}

  Since all
the limits proved above are valid for
arbitrary $x$, one
can integrate them w.r.t. $x$ within arbitrary finite interval,
say $[0,a]$

 Thus we came to the following statement.

{\bf Theorem 5} {\it In the zero bias limit $\varepsilon \to 0$:
$E = \lambda$ and
 $$
 N_{\varepsilon}(E) \,\,\,\to \,\,\,N(\lambda).
 $$
Here the integrated density of states $N_{\varepsilon}(E)$is
given by(\ref{DS}) and $N(\lambda)$ is defined by (\ref{DS0}).}


\begin{thebibliography}{99}

\bibitem{A1} L.Y.L.Shen, J.M.Rowell, Phys.Rev. {\bf 165} (1968) 566

\bibitem{A2} C.B.Duke, {\it Tunneling in Solids}, Suppl.10,
Solid State Phys. (Eds.F.Seitz, D.Turnbull) Academic, N.Y., 1969

\bibitem{A3} J.A.Appelbaum, Phys.Rev {\bf 154} (1967) 633

\bibitem{A4} N.C.Kluksdahl, A.M.Kriman, D.K.Ferry, Phys.Rev.
{\bf B39} (1989) 7720

\bibitem{A5}  G.V.Rosenblum, M.Z.Solomyak, M.A.Shubin, {\it Spectral
Theory of Differential Operators }, in Modern Topics of
Mathematics, {\bf 64}, VINITI, Moscow, 1989

\bibitem{MIR} B.S.Pavlov, G.P.Miroshnichenko, Patent
Application 5032981/25 (0113431)(Russia) from 12.03.1992

\bibitem{ANT} I.Antoniou, B.S.Pavlov, A.Yafyasov,{\it  Quantum
Electronic Devices Based om Metal-Dielectric Transition in
Low-Dimensional Quantum Structures}, in:
"Combinatorics, Complexity, Logic, Proceedings of DMTCS'96
(Eds. D.S.Bridges, C.Calude, J.Gibbons, S.Reeves, I.Witten)
Springer-Verlag, Singapore, 1996, 90-104

\bibitem{Y} A.M.Yafyasov, V.B. Bogevolnov, T.V.Rudakova,
{\it Physical Principles of Construction of Quantum Electronic
Devices Based on Metal-Dielectric Transition. Quantum
Interferentional Electronic Transistor (QIET)}, Preprint IPRT $\#$
99-95, 1995, St.Petersburg

\bibitem{D1} L.A.Dmitrieva, Yu.A.Kuperin, G.E.Rudin,
{\it Mathematical Models and Numerical Simulations for Quantum
Interferentional Device Based on Metal-Dielectric Transition},
Preprint IPRT $\#$ 170-01, 2001, St.Petersburg

\bibitem{D2} L.A.Dmitrieva, Yu.A.Kuperin, G.E.Rudin,
{\it Numerical Study of Finite SPMT Operational Regimes},
Preprint IPRT $\#$ 171-01, 2001, St.Petersburg

\bibitem{T} E.C.Titchmarsh, {\it Expansion in eigenfunctions
connected with differential operators of second order}, v.2, IL,
Moscow, 1961

\bibitem{FG} B.A.Dubrovin, V.A.Matveev and S.P.Novikov,
Rissian Math.Surveys {\bf 31}(1976) 55

\bibitem{F} N.E.Firsova,
Zap.Nauchn.Sem.Leningrad.Otdel.Mat.Inst. Steklov.(LOMI)
{\bf 51} (1975) 183

\bibitem{B1} B.S.Buslaev, L.A.Dmitrieva, Leningrad Math.J.
{\bf 1}(1990) 287

\bibitem{B2} B.S.Buslaev, L.A.Dmitrieva,
{\it Bloch Electrons in an External Electric Field}, in: {\it
Schr\"odinger Operators:
Standard and Non-Standard} (Eds. P.Seba, P.Exner), World
Scientific, Singapore, 1989, p.103

\bibitem{B3} B.S.Buslaev, L.A.Dmitrieva,
Thoer.Math.Phys.{ \bf 73} (1987) 430

\bibitem{B4} B.S.Buslaev, Russian Math.Surveys {\bf 42} (1987) 97

\bibitem{B5} B.S.Buslaev, Theor.Math.Phys. {\bf 58} (1984) 223

\bibitem{G1} V.Grecchi, A.Sacchetti, {\it Lifetime of the
Wannier-Stark resonances and perturbation theory}, preprint,
math-ph 97-211, 1997

\end{thebibliography}
\end{document}